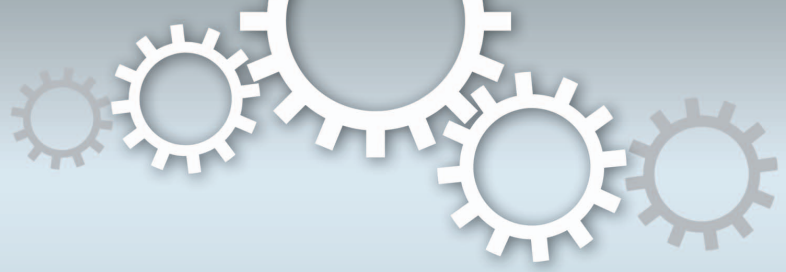

# SCIENTIFIC REPORTS

# The Increase of the Functional Entropy of the Human Brain with Age

Y. Yao[1,6]*, W. L. Lu[1,5,6]*, B. Xu[1,6]*, C. B. Li[2], C. P. Lin[3,4], D. Waxman[1] & J. F. Feng[1,5,6]

[1]Centre for Computational Systems Biology, Fudan University, Shanghai 200433, PRC, [2]Department of Biological Psychiatry, Shanghai Mental Health Centre, Shanghai Jiao Tong University School of Medicine, Shanghai 200030, PRC, [3]Department of Biomedical Imaging and Radiological Sciences, National Yang-Ming University, Taipei 11221, Taiwan, [4]Brain Connectivity Laboratory, Institute of Neuroscience, National Yang-Ming University, Taipei 11221, Taiwan, [5]Fudan University-Jinling Hospital Computational Translational Medicine Centre, Fudan University, Shanghai 200433, PRC, [6]Department of Computer Science, University of Warwick, Coventry CV4 7AL, UK.

We use entropy to characterize intrinsic ageing properties of the human brain. Analysis of fMRI data from a large dataset of individuals, using resting state BOLD signals, demonstrated that a functional entropy associated with brain activity increases with age. During an average lifespan, the entropy, which was calculated from a population of individuals, increased by approximately 0.1 bits, due to correlations in BOLD activity becoming more widely distributed. We attribute this to the number of excitatory neurons and the excitatory conductance decreasing with age. Incorporating these properties into a computational model leads to quantitatively similar results to the fMRI data. Our dataset involved males and females and we found significant differences between them. The entropy of males at birth was lower than that of females. However, the entropies of the two sexes increase at different rates, and intersect at approximately 50 years; after this age, males have a larger entropy.

T here is now a consensus that ageing is multifactorial; it is the joint outcome of genetics, the accumulation of random accidents and irreparable losses in molecular fidelity[1]. There is ample evidence that the genetic component alone plays a critical role in longevity determination[2]. This is shown in regulatory and structural changes that occur with age in miRNA[3], mRNA[4], ncRNA[5], protein expression[6] and functional MRI[7] in many species. Intuitively, these changes could be expected to correspond to changes in the functioning of the brain. But in what precise sense? This is the question we address in this work. The answer involves an explicitly quantitative way of characterizing the intrinsic ageing process of the human brain.

We used entropy to quantify the *functioning* of the brain in individuals of different ages. Accordingly, we shall describe this as the *functional entropy*.

Entropy characterizes the degree of underlying randomness of a random variable. Random variables with small entropies have a high level of predictability and hence a low level of randomness. By contrast, large entropies correspond to low levels of predictability and high levels of randomness[8].

As outlined below, we view the brain as being divided (parcellated) into a number of distinct regions. For each pair of distinct brain regions, we calculated the correlation coefficient of their neuronal activity; this characterizes the functional coupling of the two brain regions. The resulting set of correlation coefficients generates a frequency distribution. The correlation coefficient of a distinct pair of brain regions, that have been randomly selected, can be regarded as a random variable that follows this frequency distribution. We use the dispersion or variability of this random variable as a measure of the functional entropy (c.f., complexity) of the neuronal dynamics of the brain. We investigate, in this work, how this measure of the functional entropy changes with age and in Figure 1 we illustrate the behaviors of the brain's dynamics that it captures.

Figure 1 (top row) shows the situation where every brain region fluctuates over time, but is totally correlated with all other regions. In such a case, the functional entropy of correlation coefficients is zero; all correlation coefficients are unity, and hence their distribution exhibits no randomness, just predictability. A case of non-zero functional entropy occurs when a *range* of different correlation coefficients are found between different pairs of brain regions. An example of this case is given by the second row in Figure 1. See Supplementary Movies for details. In the opposite case of completely independent or incoherent activity in all regions, the correlation coefficients will all be zero and their dispersion (functional entropy) will again be zero. This means our entropy measure is sensitive to co-ordinated activity that is most interesting, namely activity that is intermediate between fully synchronised and fully incoherent brain-region dynamics.

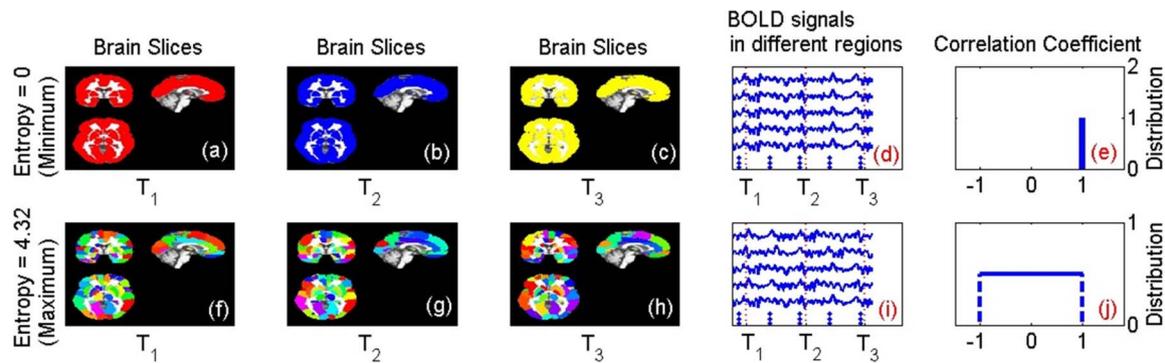

**Figure 1 | Illustration of the Functional Entropy of the Brain.** In this figure, the brain is parcellated into a number of distinct regions. Different levels of brain activity (as measured by BOLD signals) are illustrated by different colors in the brain slices of the figure. We use two artificial data sets of BOLD signals to illustrate what the functional entropy captures about brain activity. Panels (a), (b) and (c) show brain activities of brain slices at three different times ($T_1, T_2, T_3$), when the correlation coefficients between all regions of the brain are unity. In this case, all regions of the brain have the same color since they are behaving synchronously. Panel (d) shows the BOLD signals in different brain regions, for this case. Panel (e) shows the corresponding distribution of the correlation coefficients (a 'spike' located at a correlation coefficient of unity). The functional entropy for this case is zero (the minimum possible value). Panels (f), (g) and (h) show the brain activities at three different times ($T_1, T_2, T_3$), when all correlation coefficients are generally different (so all regions have different colors, indicating that all regions are behaving asynchronously). Panel (i) shows BOLD signals in different brain regions, for this case. Panel (j) shows the corresponding distribution of the correlation coefficients (a uniform distribution). The functional entropy in this case is 4.32 bits (the maximum possible value). See Supplementary Movies for details.

The functional entropy effectively measures the dispersion (or spread) of functional connectivities that exist within the brain. We initiated this study under the assumption that the dispersion of functional connectivities is related to the age of the brain.

For the current paper, we collected fMRI data from 1248 individuals, ranging from 6 to 76 years of age. This provided a unique opportunity to characterize the ageing process of the human brain.

In the analysis of our fMRI dataset of differently aged individuals, we found that, at the population level, the functional entropy of the human brain (as calculated below) has a definite tendency to increase over time. This can be viewed as there being a higher level of randomness in the way different brain-regions functionally interact with one another.

Beyond showing that the functional entropy of the brain has the tendency to increase with age, we find quantitative differences between the entropies of males and females. In newborn males the functional entropy has a mean value of 3.536 bits; it is approximately 0.06% larger in newborn females, with a value of 3.555 bits. However, there are is also a difference in the *rate* at which the functional entropies change in the two sexes. In males, the functional entropy increases at a mean rate of approximately 0.0015 bits/year but in females it increases at the slower mean rate of 0.0011 bits/year. The different values of the functional entropy in newborns, and the different rates of increase in the two sexes, lead to the entropies of the two sexes approaching one another and then crossing. This crossover in entropies occurs at approximately 50 years of age. Beyond this age, the pattern of entropies exhibited at birth is reversed, with males then having the larger functional entropy.

Given that the current world life expectancy is 65.59 years in males and 69.73 years in females[9], we estimate that at these ages, the functional entropy of the brain will be 3.633 bits in males and 3.647 bits in females. Thus the mean functional entropy change, from birth to life expectancy, is 0.097 bits in males, and 0.092 bits in females, even though female life expectancy is higher.

In addition to determining the functional entropy of the *whole brain*, we have determined the functional entropy of different *regions* of the brain, again using correlations between the neuronal activity of different brain regions. We find that different brain regions have different entropic characteristics. Typically, the observed changes are monotonic, but not all brain regions have increasing entropies. There are some regions where the functional entropy increases, others where it decreases, and a third set where the functional entropy remains almost constant. With L and R denoting left and right brain regions, we find that the brain regions with the *fastest* rate of functional entropy increase are: the L and R paracentral lobules, the R olfactory cortex, the L middle frontal area, the L and R hippocampi and the L and R parahippocampal gyrus. We note that the hippocampus is well known to be associated with both short and long term memory formation[10]. By contrast, the L and R insulars represent regions whose entropies most rapidly *decrease* with the age of an individual. Clear changes in regions of the brain, with age, are also found in other studies, INS[11], PCL[12], OLF[13], MFG[14], HIP[15] and PHG[16]. We note that the functional entropy of the whole brain is not simply an average over entropies of individual brain regions. Accordingly, our findings, that the functional entropy of the whole brain increases with age while some regions of the brain exhibit decreasing entropies, are compatible.

We have used a computational model[17] based on diffusion tensor imaging (DTI) data to investigate the origins of the relationship between functional entropy and age. Extensive experimental data indicates that there is significant loss of neuron number with age, and this is accompanied by the excitatory receptor number (especially NMDA) decreasing with age[18]. Our computational model (see Supplementary Information for details) yields a brain entropy that decreases when the excitatory connection strength and neuron number in each brain region are simultaneously reduced.

To motivate the definition of functional entropy that we use in this work, let us consider the following example of an analysis we carried out.

We parcellated the whole brain of three individuals into 90 regions, based on the AAL atlas[19]. These were healthy males aged 24, 49 and 69 years, which we describe as 'young', 'middle-aged' and 'elderly'. For these, we calculated the correlation coefficient between the BOLD signal of the thalamus in the right hemisphere and each of 45 brain regions in the left hemisphere (see Supplementary Information, Table S2). These signals are represented in the left-hand of panels (a), (b) and (c) of Figure 2. Differences between the three individuals show up which are found in more extensive analyses.

The distribution of the correlation coefficients of the elderly male (red histogram in Figure 2d) is more widely spread than that of the young male (blue histogram) and middle-aged male (white histogram). This leads to the elderly male having a larger functional entropy than that of the middle aged male, who has a yet larger functional entropy than that of the young male (see Figure 2d). This implies that the dispersion of correlations, between the right thalamus and a region of the brain in the left hemisphere, is typically an increasing function of age. This conclusion is found to hold in a full analysis, where a pairwise comparison of all regions in the brain is used, rather than just comparing regions in the left hemisphere with the right hemispheric thalamus.

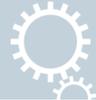

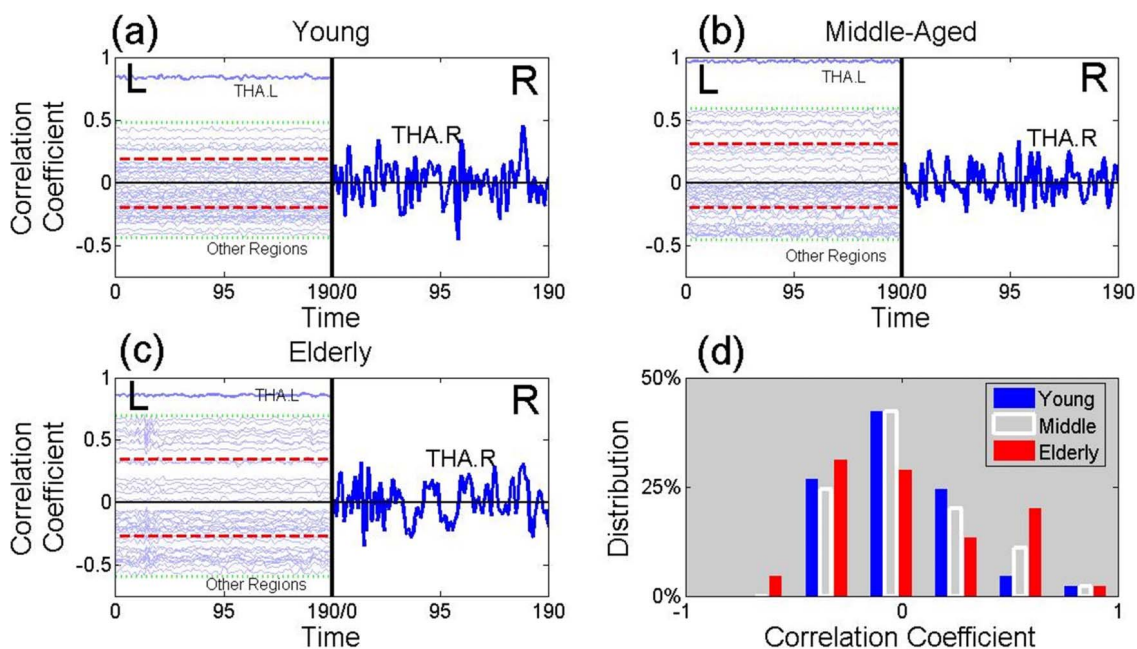

**Figure 2 | The Origin of the Functional Entropy.** This figure presents time series from BOLD signals. The left-hand sides of panels (a), (b) and (c) contain 45 time series from brain regions of the left hemisphere. The vertical location of a time series, from a given brain region, is given by value of the correlation coefficient of that region with the right thalamus. Panel (a) is from a healthy young male (age 24 years), panel (b) is from a healthy middle-aged male (age 49 years) and panel (c) is from a healthy elderly male (age 69 years). The two horizontal red lines in panels (a), (b) and (c) give, separately, the mean over either positive or negative correlation coefficients. Thus the separation of these lines is a measure of the width of the distribution. In the right halves of panels (a), (b) and (c), the time series of the right thalamus is plotted (using a different vertical scale). Panel (d) gives a histogram of the correlation coefficients of the young male (in blue), the middle-aged male (in white) and the elderly male (in red).

## Results

We have carried out a full analysis to determine the functional entropy in all individuals in our data set. Figure 3a shows the functional entropy of the full data set as a function of age, without taking into account gender differences. In Figure 3b we present a running average of the functional entropy for males and females, with an averaging window of 25 years, whose choice is a compromise between stability and being substantially smaller than the maximum age; choosing windows from 19 to 30 years does not significantly affect any conclusions we draw (see Supplementary Information 6). In Figures 3c and 3d, we give the results for the functional entropy in males and females separately, which are different at birth (3.5336 bits in males, 3.5547 bits in females) and which have different rates of change (0.0015 bits/year in males, 0.0011 bits/year in females). The Pearson correlation between entropy and age is strongly significant ($r = 0.23$, $N = 610$, $p = 5.6 \times 10^{-9}$ for males and $r = 0.15$, $N = 634$, $p = 1.51 \times 10^{-4}$ for females) These lead to the crossover that can be seen in Figure 3b, at an approximate age of 50 years, and will be considered in the Discussion.

For each brain region considered in this study, we have also determined trends in their individual entropies over time. The functional entropy of a given region is determined from the set of correlation coefficients between that region and the remainder of the brain. In Figure 4a we have used color to show the trend: the warmer (redder) the color, the more positive the rate of increase. Figure 4a shows that the frontal area is more likely to have a higher rate of increase in functional entropy with age than an average region. By contrast, the occipital area exhibits an apparent resistance to functional entropy change, and remains largely unaltered over time. This pattern is considered in detail in the Discussion.

The brain regions that exhibit the most significant ($p < 1 \times 10^{-7}$) functional entropy changes with age are shown in Figure 4e, and indicated spatially in Figure 4b, c and d. The results are based on the slope versus age, excluding any effects of gender. The most significant increasing or decreasing regions with ageing have already been shown in the Introduction. They are INS ($r = 0.157$, $p = 2.47 \times 10^{-8}$), PCL ($r = 0.217$, $p = 1.07 \times 10^{-14}$), OLF ($r = 0.269$, $p < 1 \times 10^{-15}$), MFG ($r = 0.198$, $p = 1.56 \times 10^{-12}$), HIP ($r = 0.221$, $p = 3.11 \times 10^{-15}$) and PHG ($r = 0.224$, $p = 1.33 \times 10^{-15}$) ($N = 1246$), which will be considered in detail in the Discussion.

To investigate whether functional entropy can theoretically increase with age, a computational model from Deco et al.'s work[17] was employed. In essence, these authors formulated a detailed model of a brain network as a global random attractor. This offers a realistic mechanistic model, at the level of each single brain area, which is based on spiking neurons and realistic AMPA, NMDA, and GABA synapses. The global architecture of the model is shown in Figure 5a. After obtaining the neuronal dynamics in each brain region, the fMRI BOLD signal was simulated by means of the Balloon-Windkessel hemodynamic model[20]. We found that the functional entropy increases with decreasing excitatory connection strength (AMPA and NMDA) within the excitatory population of each brain region, shown in Figure 5b. To match functional entropy of the human data, we determined a reliable range of connection strengths [1.78, 1.81] (indicated by two red dashed lines in Figure 5b) by comparing the two least-square lines in Figure 3a and Figure 5b. Figures 5c and 5d illustrate the simulated BOLD signal with two different connection strengths; the distribution of the correlation coefficients is broader for the smaller connection strength. Specifically, Figure 5c, with a larger connection strength is similar to the result for young people in Figure 2a, while Figure 5d, with a smaller connection strength, is similar to that of elderly people in Figure 2c. Additionally, assuming a positive correlation between excitatory neuron number and connection strength (shown in Supplemental Figure S4), we find that the excitatory neuron number in each brain region decreases from 1130 to 888 (approximately a 20% loss). It is know from experiments that both the neuron number and excitatory receptors number in the human brain reduce with age. Thus we conclude that this simplified computational model does lead to a relationship between functional entropy and age that is similar to that observed.

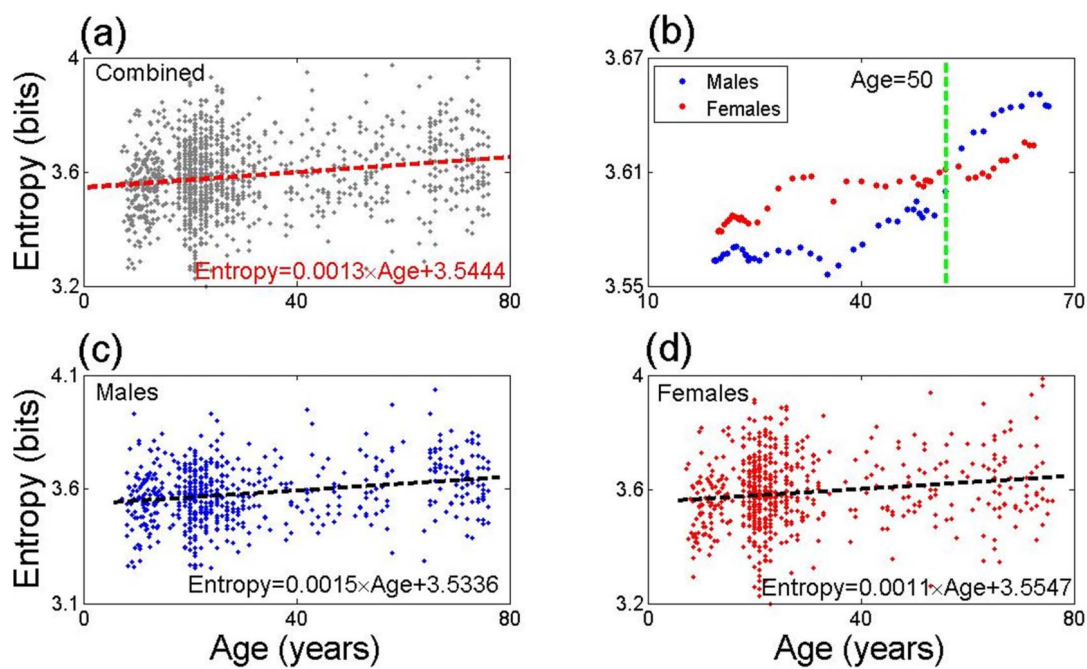

**Figure 3 | Functional Entropy vs. Age.** Panel (a) is a plot of the functional entropy of individuals versus their age (pooling results from males and females). A mean rate of increase of the entropy of 0.0013 bits/year was found from the data. Panel (b) contains a plot of the running average of the entropy, versus age, with a window of width of 25 years adopted. There is a crossover in the male/female entropies in the vicinity of 50 years of age. Panels (c) and (d) plot the entropy versus age of males and females. Males have a lower initial value of the entropy than females, but a faster mean rate of increase. The linear correlation between entropy and age is strongly significant ($p = 5.6 \times 10^{-9}$ for males and $p = 1.51 \times 10^{-4}$ for females, r equals 0.23 for males and 0.15 for females, degree of freedom of 610 and 634, respectively.)

## Discussion

In this paper, we have calculated the functional entropy from correlations between BOLD signals of different regions of the human brain. We have found that the functional entropy is an increasing function of age. Our approach allows a novel explicitly quantitative characterization of the functional entropy in the human brain, and the results obtained are closely consistent with a computational model where ageing is incorporated by decreasing the NMDA conductance and the number of excitatory neurons.

The functional entropy discussed in the present work is estimated only for the resting state. The value of the functional entropy determined does not directly apply to task driven fMRI, since external tasks result in activation of brain circuits specific to the task at hand. These are likely to confound any analysis along the lines presented here, which aims to capture genuinely *intrinsic* aspects of the human brain.

Our results are in agreement with well-known facts in neuroscience. The histogram of correlation coefficients of the functional network, as plotted in Figure 3d, is more peaked in schizophrenic patients, around zero, and hence generally leads a smaller functional entropy than would be expected for a patient's age. (This conclusion is validated in the Supplementary Information, Section 7). This feature closely corresponds to the disconnection hypothesis[21] of schizophrenics, which has been established at an early point in the literature.

Consider, next, the pattern of increase of functional entropy shown in Figure 4a. Most brain regions show a trend of the functional entropy increasing with age. This fits with the idea about the hemispheric asymmetry reduction in older adults[22] (HAROLD): the increasing functional entropy comes from correlation coefficients moving away from zero, when the age is increased. This also corresponds to the left and right hemispheres becoming more symmetric in activity with age, in the resting state. In addition to its similarity to HAROLD, this pattern is also similar to the Posterior-Anterior Shift in Aging[23] (PASA), which is an age-related ascension in frontal activity coupled with comparatively reduced occipital activity, as the functional entropy change speed in the frontal area is much faster than that in the occipital area in our results. Moreover, Figure 2 also partly reflects HAROLD, since the correlation coefficients of the links connecting the right thalamus and the left brain regions move away from zero with ageing, which leads to the increase symmetry in the inter hemisphere.

A robust feature of the analysis presented here concerns the crossover that is observed, around 50 years of age (50.2, exactly), between the mean entropies if male and female brains of healthy individuals (see Figure 3b). Below the crossover age, males have a lower mean functional entropy. Above this age, the roles are reversed and females have a lower mean functional entropy. Physical differences of the brains of males and females may account for the functional entropy differences at birth, and different hormonal environments of male and female brains throughout life could play a role in the different rates of change of the functional entropy throughout life. There are a number of phenomena which occur in the vicinity of an age of 50 years, where the crossover in entropies occurs. (i) The mean age of occurrence of the female menopause is close to 50 years. (ii) A cross-over occurs, around 50 years, in the increase of world knowledge and the decreasing speed of processing, working memory and long-term memory[24]. (iii) It has been recognized that In addition, the unhappiness of human beings peaks around 50 years; people of that age have the most chance to become depressed[25].

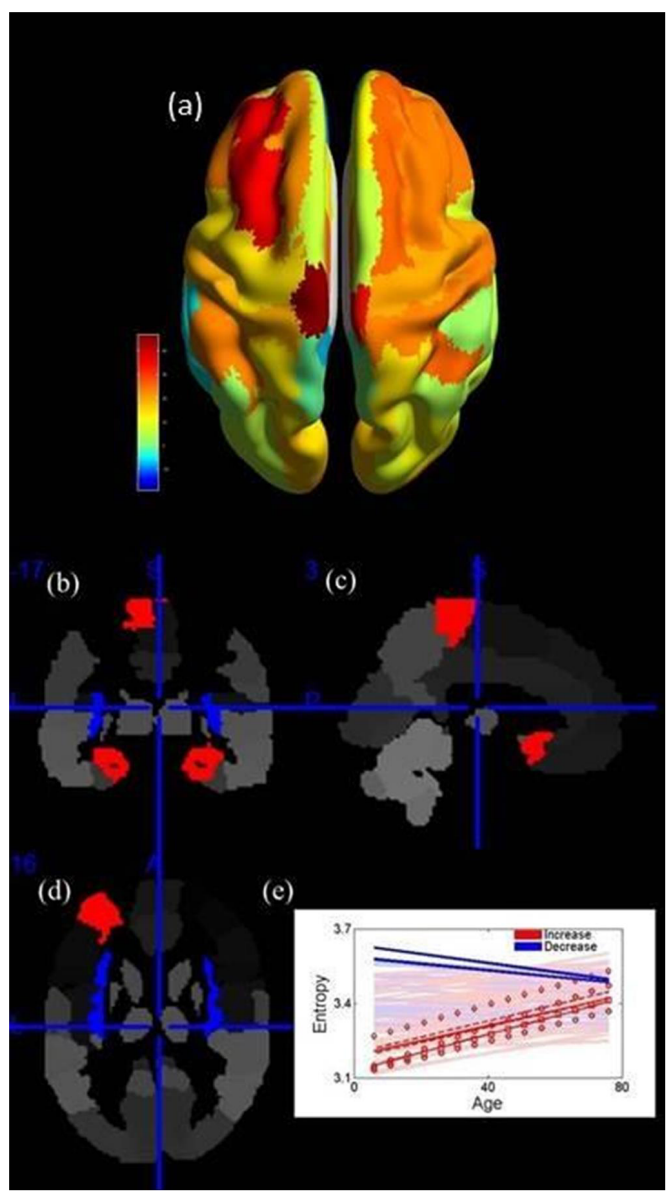

**Figure 4 | Functional Entropy of Brain Regions.** Panel a shows the trend exhibited by the functional entropy of each brain region with age. The warmer (redder) the color, the larger the rate of change of the entropy. Panels (b), (c) and (d) give the coronal, sagittal and axial views of the brain, and show the brain regions with the most rapidly changing entropies. Red ones show the brain regions with the most rapidly increasing entropy, blue ones show the most rapidly decreasing regions. Panel (e) contains the trends of all the brain regions with significant ones being in the brightest colors. In panel (e), bright blue lines show the behavior of the insula (L, R); bright dotted red lines indicate the paracentral lobule (L, R); solid red lines indicate the hippocampus (L, R); dashed red lines indicate the parahippocampal gyrus (L, R); red squares represent the olfactory cortex (R); diamonds represent the middle frontal gyrus (L).

Concerning the rates at which functional entropy changes, we note that in the Figure 3, the increasing rate of the functional entropy of females is lower than that of males. This may be related to the fact that estrogen protects women's brains[26], which decreases the risk of Parkinson's[26] and Alzheimer's diseases[27].

In the Results, we find that INS, PCL, OLF, MFG, HIP and PHG exhibit the most significant rates of change with age. The six brain regions have extensive references supporting their relation with ageing. For example, INS and PHG represent significant volume atrophy with ageing[11,16]. HIP holds a significant neuron decrease in some parts[15]. Moreover, PCL, OLF and MFG show more activations in elderly people compared with young[12–14]. Functionally, HIP and PHG have a high relationship with memory[28] and OLF connects with smell[29], which is typically deficient in the elderly[30]. In addition, PCL and MFG are responsible for the processing of higher information[31]. The various results found in other studies validate our results to some sense. Furthermore, the pattern, that INS exhibits a significantly decrease of functional entropy, shows that the insula is quite special since the total functional entropy grows higher with age. Although we have no knowledge why the INS exhibits this kind of pattern, this deserves further investigation and may relate to the various functions of INS.

There are a number of sources of error and noise which might influence the robustness of the results. We note that despite the size of the data set, it was necessary to average over a fairly extensive time window, to see a general trend in the functional entropy. While such an averaging is fully justified, to eliminate noise in the data, it also has the effect of reducing or eliminating features in the data which occur on timescales that are smaller than the averaging window. We have considered the influence of the atlas and found that the results are robust with respect this (see Supplementary Information, Section 16).

The age distribution of the subjects was not uniform (See Figure S11). About half of the subjects were aged in their twenties when scanned. To determine if the influence of this non uniform age distribution affected the correlation found between entropy and age, we conducted resampling by selecting six subjects from each age with equal probability and calculated the correlation between functional entropy and age. We repeated this resampling nine times and found that every trial showed a significant positive correlation between functional entropy and age (See Figure S21). From this, we conclude that the correlation between functional entropy and age is not an artifact of the non-uniform age distribution.

Other influences, for example the pre-processing have not been considered since they are standard and well-accepted.

In contrast to the existing definitions of entropies to measure the randomness of brain activities in time-domain[32,33], our assessment of complexity was based upon functional connectivity matrices or correlations of these brain activities. As shown in the literature[34], the randomness of brain activity decreases with age, which implies a decrease of the entropy in time domain. However, we found, the functional entropy increases with age. The relationship between these two different sorts of entropy and the different trends they exhibit with age will be investigated in the near future.

Although functional connectivity provides a simple and direct way of assessing the statistics (complexity) of neuronal fluctuations, there is an important qualification that we have to make: in basing our analyses on observed hemodynamic fluctuations, we assume that the underlying changes can be ascribed to neuronal fluctuations. This means that we are implicitly assuming that there are no age-related changes in neurovascular coupling or the measurement of hemodynamics (for example, no differences in head movement). Either of these differences could affect the correlations and therefore age-related changes in their characterisation. Future work could use dynamic causal modelling to separately estimate the neuronal complexity, while allowing for subject-specific and age-related changes in hemodynamic parameters.

Finally, let us point out that our data is a cross sectional study and hence a repetition of the analysis, in the form of a longitudinal study, is warranted. However, due to the small rate of increase of the functional entropy, a significant increase in the functional entropy is only likely to be weakly observed over a large time span. Accordingly, it might be more promising to carry out a longitudinal study in other models, for example, in monkeys.

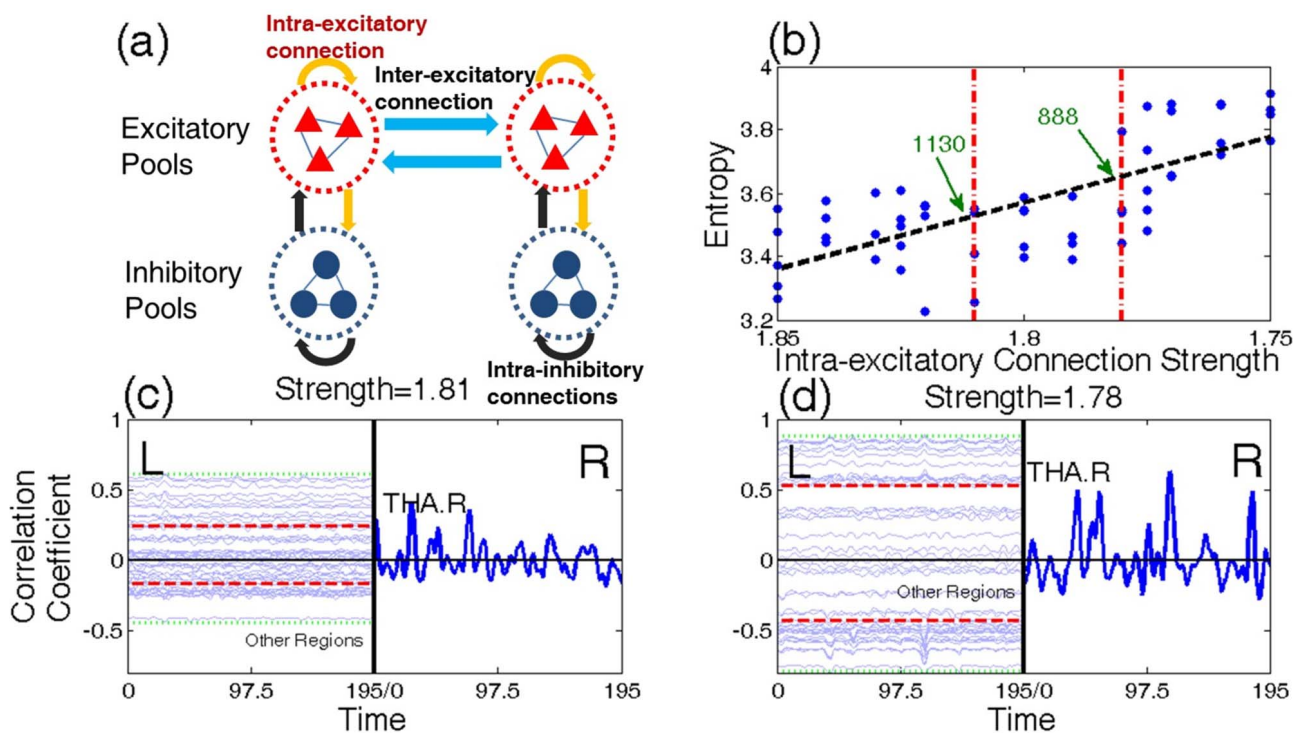

**Figure 5 | Computational Model.** Panel (a): Schematic representation of brain network. Each brain region is comprised of excitatory pyramidal cells (red triangles) and inhibitory interneurons (blue circles). Yellow and black arrows describe excitatory and inhibitory connections between neurons in each brain region respectively, and blue arrows show excitatory connections between neurons in different brain regions. Panel (b): Functional entropy versus intra-excitatory connection strength. The black dash line represents the least-square line of blue dots (different trials), and the linear correlation between functional entropy and connection strength is statistically significant ($1 \times 10^{-10}$). The two red dashed lines show a range of connection strengths, [1.78, 1.81], which make the corresponding functional entropy match the human data. From the relationship between strength and excitatory neuron number (see Supplementary Information Figure S4), the neuron number range is limited to [888, 1130] (indicated by green arrows). Panel (c) and (d): Time series of simulated BOLD signal. Similar to Figure 1, the left hand sides of panels (c) and (d) contain 45 time series arising from brain regions in the left hemisphere. The time series are vertically located, according to their phase difference with the right thalamus. The corresponding connection strength in (c) and (d) is 1.81 and 1.78, respectively. (c) is similar to the result of young people and (d) is similar to that of elderly people.

## Methods

**Subjects.** Our meta-analysis included 26 datasets, with a total of 1248 samples. These covered a range of individuals from 6 to 76 years of age. We excluded samples of poor quality.

Note that 22 of these datasets came from the 1000 Functional Connectomes Project (http://fcon_1000.projects.nitrc.org/), where data came from all over the world including China, Britain and United States. In these 22 data sets, there were 842 samples, with ages ranging from 18 to 73 years, with 357 of them male. The mean age was $28.3 \pm 12.3$ years. Details are listed in Supplementary Table S1. The fMRI scans performed by 1000 Functional Connectomes Project were carried out in accordance with the guidelines issued by the local ethical committees of the various research institutes, which can be found in their website. And informed consent was obtained from all subjects.

Another 2 datasets were from the ADHD-200 Consortium for the global competition (http://fcon_1000.projects.nitrc.org/indi/adhd200/). One of these was from the Phyllis Green and Randolph Cowen Institute for Pediatric Neuroscience at the Child Study Center, New York University Langone Medical Center, New York, New York and the Nathan Kline Institute for Psychiatric Research, Orangeburg, NY, USA. The other came from the Institute of Mental Health, Peking University and the National Key Laboratory of Cognitive Neuroscience and Learning, Beijing University. Since they were both concerned about ADHD classification, their data were ADHD patients versus controls (normal individuals). We cannot apply the data from individuals with ADHD disorder, since their brains may be different from normal people, and we used only the controls, namely the normal people from the 2 datasets. In total, there were 241 samples, with 134 of them male. Their mean age was $11.83 \pm 2.47$ years, so they were teenagers/children. Since most samples in the other dataset were more than 18 years old, we selected the two datasets to complete our story about ageing in the range of 6 years to 18 years old. The fMRI study in Peking University was approved by the Research Ethics Review Board of Institute of Mental Health, Peking University. Informed consent was also obtained from the parent of each subject and all of the children agreed to participate in the study. The fMRI scans were carried out in accordance with the guidelines issued by the local ethical committee, and informed consent was obtained from all subjects.

There was also one dataset of elderly people, covering 117 samples from the Department of Psychiatry, Tongji Hospital, Tongji University School of Medicine and Department of Biological Psychiatry, Shanghai Mental Health Centre, Shanghai Jiao Tong University School of Medicine. In this dataset, 73 were male. The mean age was $70.42 \pm 3.52$ years. With a designed health status checklist, we excluded individuals with: obvious cognitive decline, a diagnosis of AD, serious functional decline (having difficulty with independent living), as well as individuals with major medical or psychiatric conditions such as cancer, current chemotherapy/radiation treatment, major depression, and schizophrenia. This study was approved by the Human Research Ethics Board of Tongji Hospital in Shanghai, China and all participants gave written informed consent before being enrolled in the study.

The last dataset was from the Department of Biomedical Imaging and Radiological Sciences, National Yang-Ming University, Taipei, Taiwan and the Brain Connectivity Laboratory, Institute of Neuroscience, National Yang-Ming University, Taipei, Taiwan. There were 48 samples in the dataset. All were male and covered a range of ages from 21 to 76 years. The mean age was $43.8 \pm 17.0$ years and all individuals were normal and healthy. Moreover, the T1-image and DTI data of these 48 samples were also applied in our paper. The fMRI scans from National Yang-Ming University were carried out in accordance with the guidelines issued by the local ethical committee, and informed consent was obtained from all subjects.

**Data acquisition.** The various ways that data were acquired in the 22 datasets from the 1000 Functional Connectomes project can be found in their website, http://fcon_1000.projects.nitrc.org/. Moreover, the acquisition of the two datasets from the ADHD-200 Consortium for the global competition was in their website, http://fcon_1000.projects.nitrc.org/indi/adhd200/.

In the dataset from Shanghai, all people underwent functional scanning using a Siemens Trio 3 T scanner at East China Normal University, Shanghai, China. Foam padding was used to minimize head motion for all subjects. Functional images were acquired using a single-shot, gradient-recalled echo planar imaging sequence (repetition time = 2000 ms, echo time = 25 ms and flip angle = 90 degrees). Thirty-two transverse slices (field of view = 240 × 240 mm², in-plane matrix = 64 × 64, slice thickness = 5 mm, voxel size = 3.75 × 3.75 × 5 mm³), aligned along the anterior commissure–posterior commissure line were acquired. For each subject, a total of 155 volumes were acquired, resulting in a total scan time of 310 s. Subjects were instructed simply to rest with their eyes closed, not to think of anything in particular, and not to fall asleep.

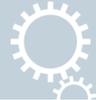

At last, in the dataset from Taiwan, all people underwent structural, functional and diffusion tensor imaging scanning using a Siemens Trio 3 T scanner at National Yang-Ming University, Taiwan. Foam padding was used to minimize head motion for all subjects. Functional images were acquired using a single-shot, gradient-recalled echo planar imaging sequence (repetition time = 2500 ms, echo time = 27 ms and flip angle = 77). Fourty-three transverse slices (field of view = 220 × 220 mm², in-plane matrix = 64 × 64, slice thickness = 3.4 mm, voxel size = 3.44 × 3.44 × 3.4 mm³), aligned along the anterior commissure–posterior commissure line were acquired. For each subject, a total of 200 volumes were acquired, resulting in a total scan time of 500 s. Subjects were instructed simply to rest with their eyes closed, not to think of anything in particular, and not to fall asleep. The diffusion tensor images covering the whole brain were obtained using spin echo-based diffusion tensor imaging sequence, including 30 volumes with diffusion gradients applied along 30 non-collinear directions (b = 1000 s/mm²) and three volumes without diffusion weighting (b = 0 s/mm²). Each volume consisted of 63 contiguous axial slices (repetition time = 11000 ms, echo time = 104 ms, flip angle = 90 degrees, field of view = 100 × 100 mm², matrix size = 128 × 128, voxel size = 2 × 2 × 2 mm³). To improve the signal to noise ratio, the entire sequence was repeated three times. Subsequently, high-resolution T1-weighted anatomical images were acquired in the sagittal orientation using a magnetization-prepared rapid gradient-echo sequence (repetition time = 3500 ms, echo time = 3.5 ms, flip angle = 7, field of view = 256 × 256 mm², matrix size = 256 × 256, slice thickness = 1 mm, voxel size = 1 × 1 × 1 mm³ and 192 slices) on each subject.

**Data preprocessing.** Firstly, we dealt with the two datasets from the controls of ADHD 200 Consortium. Before functional image preprocessing, the first four volumes were discarded, to allow for scanner stabilization. Briefly, the remaining functional scans were first corrected for within-scan acquisition time differences between slices, and are then realigned to the middle volume, to correct for inter-scan head motions. The functional scans were then spatially normalized to a standard template[35] (Montreal Neurological Institute) and resampled at 4 mm × 4 mm × 4 mm voxel resolution. After normalization, the Blood Oxygenation Level Dependent (BOLD) signal of each voxel was first detrended to remove any linear trend and then passed through a bandpass filter (0.009 Hz < f < 0.08 Hz) to reduce low-frequency drift and high-frequency physiological noise. Finally, nuisance covariates including head motions, global mean signals, white matter signals, and cerebrospinal fluid signals were regressed out. An automated anatomical labeling (AAL) atlas[19] was used to parcellate the brain into 90 regions of interest (ROIs), with 45 in each hemisphere. The names of the ROIs and their corresponding abbreviations are listed in supplementary Table S2. We thank Carlton Chu, Virginia Tech's ARC, the ADHD-200 consortium, and the Neuro Bureau for what they have done for us.

Let us now consider all other datasets. The first 10 volumes of these datasets were discarded, to allow for scanner stabilization and the subjects' adaptation to the environment. fMRI data preprocessing was then conducted by Statistical Parametric Mapping[36] (SPM8) and a Data Processing Assistant for Resting-State fMRI[37] (DPARSF). The remaining functional scans were first corrected for within-scan acquisition time differences, between slices, and then realigned to the middle volume, to correct for inter-scan head motions. Subsequently, the functional scans were spatially normalized to a standard template[35] (Montreal Neurological Institute) and resampled to 3 × 3 × 3 mm³. Data was then smoothed, and after normalization and smoothing, BOLD signals of each voxel were first detrended, to remove any linear trend, and then passed through a band-pass filter (0.01–0.08 Hz) to reduce low-frequency drift and high-frequency physiological noise. Finally, nuisance covariates including head motions, global mean signals, white matter signals and cerebrospinal signals were regressed out from the BOLD signals. After data preprocessing, the time series were extracted in each ROI by averaging the signals of all voxels within that region and then linearly regressing out the influence of head motion and global signals. In our present study, the automated anatomical labeling atlas[19] (AAL) was used to parcellate the brain into 90 regions of interest (ROIs) (45 per hemisphere). The names of the ROIs and their corresponding abbreviations are listed in supplementary Table S2.

On the other hand, we also applied the DTI data of the 48 samples from Taiwan to construct the connection matrix in our neuron model. We used the FSL[38] to remove the eddy-current and extract the brain mask of B0 image. Then, we used the TrackVis[39] (http://www.trackvis.org/) to obtain the fiber images by the deterministic tracking method. After that, we used the T1 data to extract the native template for every person by FSL. Thus, we could count the number of fibers connecting every two brain regions. All the processes were performed by a pipeline named PANDA[40] from Beijing Normal University.

**Entropy calculation method.** After data preprocessing, the time series were extracted in each ROI by averaging the signals of all voxels within the region. The 90 regions were based on a selected atlas, say the AAL Template. After that, we calculated the Pearson Correlation Coefficient of every pair of regions. Since our atlas was the AAL template, we had 4005 function links connecting every two regions. Thus we constructed a whole brain functional network.

Given 4005 different correlation numbers, we required an indicator to represent features of the whole-brain functional network, and thus considered two possible approaches.

1) In the first approach, we considered the values of the correlation coefficients as a realization of a random variable. The range of this was [−1, 1]. We then defined the brain functional entropy as the relative entropy, i.e., the KL (Kullback-Leibler) divergence[41] from the correlation distribution to a reference Lebesgue measure. In practice, we did not have a continuous distribution of correlation coefficients, but 4005 correlations values from each individual. We thus separated all 4005 realizations into 20 class intervals of equal width, and determined the frequency ($p_i$) of each class. These frequencies were used to calculate the Shannon entropy (sum of $-p_i*\log(p_i)$)[42] of the whole brain. This can be considered as the functional entropy of a discrete distribution[42]. The justification for replacing the relative entropy by the entropy of a discrete distribution are given in the Supplementary Information.

2) The alternative approach was to calculate the functional entropy for every single brain region. That was, consider the functional links between the selected region and the other 89 regions. Thus, we could extract 89 correlations. We also separated these into 20 equal-width class intervals, and then calculated the Shannon entropy of each region. It should be emphasized that the functional entropy of the whole brain is not simply an average over entropies of individual brain regions (see the Supplementary Information, Section 3.2 for more details).

Moreover, the base of the logarithm in the paper is 2. Thus, the unit of the functional entropy in the paper is the bit (binary digit).

**Statistical inference.** In this paper, several statistical inference methods were used. The first and most basic one was Pearson product-moment correlation coefficient[43]. We used it to describe the relationship between every pair of brain regions and the connection between the functional entropy and ageing. In addition, we also used partial correlations[43]. Moreover, we applied a linear regression analysis[44] to determine any trends of the functional entropy of normal people, with age.

**Computational model.** We used a computational model to simulate BOLD signals. Every brain region served as a node in a large scale network, which consisted of a population of excitatory pyramidal neurons and a population of GABAergic inhibitory neurons, which were all-to-all connected. The communication between every two nodes was through synaptic connections between excitatory neurons in those areas. Those inter-regional connections were defined by number of fibers linking different regions in a structural connectivity matrix and by a connection strength. The neuroanatomical matrix was obtained by Diffusion Tensor Imaging. Here, we used averaged structural matrix from 46 healthy people, which is shown in Figure S13. In addition, we considered a group of single neurons for each node in a global network. Specifically, we applied the biophysically realistic attractor network model consisting of integrate-and-fire spiking neurons with excitatory (AMPA and NMDA) and inhibitory (GABA-A) synaptic receptor types in the microscopic level. The spiking activity of neurons was characterized by the dynamics of the membrane potential. We can use this model to simulate BOLD signals. This then allowed the calculation of the functional entropy for different parameter values and the extraction of the relationship between trends in the functional entropy and parameters, such as the intra-connection in every brain region. Details are given in the Supplementary Information.

## Acknowledgements

YY and BX are supported by the China Scholarship Council (CSC). JF is a Royal Society Wolfson Research Merit Award holder, partially supported by National Centre for Mathematics, Interdisciplinary Sciences (NCMIS) of the Chinese Academy of Sciences and the Key Program of National Natural Science Foundation of China (No. 91230201). WL is jointly supported by the Marie Curie International Incoming Fellowship from the European Commission (FP7-PEOPLE-2011-IIF-302421), the National Natural Sciences Foundation of China (No. 61273309), the Foundation for the Author of National Excellent Doctoral Dissertation of PR China (No. 200921), Shanghai Rising-Star Program (No. 11QA1400400), and also by the Laboratory of Mathematics for Nonlinear Science, Fudan University. CBL is supported by National Natural Science Foundation of China (No. 30770769). CPL is supported by grant NSC 101-2911-I-008-001 (the Center for Dynamical Biomarkers and Translational Medicine, National Central University, Taiwan).

## Author contributions

J.F.F. formulated the idea of the paper and supervised the research. C.B.L. and C.P.L. collected some of the MRI data samples. They also made useful suggestions for the study. Y.Y. preprocessed and analyzed the data. W.L.L. helped with data analysis. W.L.L. and Y.Y. also carried out the mathematical part of this study. B.X. constructed the computational model. D.W. summarized the research and improved the study. He also strengthened the meaning of the entropy in this investigation. Y.Y., W.L.L., B.X., D.W. and J.F.F. wrote the manuscript.

## Additional information

**Supplementary information** accompanies this paper at http://www.nature.com/scientificreports

**Competing financial interests:** The authors declare no competing financial interests.

**How to cite this article:** Yao, Y. *et al.* The Increase of the Functional Entropy of the Human Brain with Age. *Sci. Rep.* **3**, 2853; DOI:10.1038/srep02853 (2013).

8